\documentclass{elsart}
\newcommand{\vecbm}[1]{\mbox{\boldmath#1}}
\newcommand{\nvec}[1]{\stackrel{\rightarrow}{#1}}
\newcommand{\lora} {\boldmath$\longrightarrow$}
\usepackage{natbib}

 \usepackage{graphicx}

\begin{document}

\begin{frontmatter}



\title{Reconciliation of Statistical Mechanics \\ and Astro-Physical Statistics.\\
The errors of conventional \underline{canonical} Thermostatistics}
Presented at the NBS2005 at the Observatoire de Paris

\author{D.H.E.Gross}

\address{Hahn-Meitner-Institut Berlin,\\
    and Fachbereich Physik der Freien Universit\"at, Berlin, Germany}
\ead{gross@hmi.de, WEB: http://www.hmi.de/people/gross}

\begin{abstract}Conventional thermo-statistics address infinite homogeneous systems within
the canonical ensemble. (Only in this case this is equivalent to
the fundamental microcanonical ensemble.) However, some 170 years
ago the original motivation of thermodynamics was the description
of steam engines, i.e. boiling water. Its essential physics is the
separation of the gas phase from the liquid. Of course, boiling
water is inhomogeneous and as such cannot be treated by
conventional thermo-statistics. Then it is not astonishing, that a
phase transition of first order is signaled canonically by a
Yang-Lee singularity. {\em Thus it is only treated correctly by
microcanonical Boltzmann-Planck statistics}. It turns out that the
Boltzmann-Planck statistics is much richer and gives fundamental
insight into statistical mechanics and especially into entropy.
This can be done to a far extend rigorously and analytically. As
no extensivity, no thermodynamic limit, no concavity, no
homogeneity is needed, it also applies to astro-physical systems.
The deep and essential difference between ``extensive'' and
``intensive'' control parameters, i.e. microcanonical and
canonical statistics, is exemplified by rotating, self-gravitating
systems. In the present paper the necessary appearance of a convex
entropy $S(E)$ and negative heat capacity at phase separation in
small as well macroscopic systems independently of the range of
the force is pointed out. Thus the old puzzle of stellar
statistics is finally solved, the appearance of negative heat
capacity which is forbidden and cannot appear in the canonical
formalism.
\end{abstract}


\end{frontmatter}

\section{Introduction}\label{introduction}

Conventional statistical mechanics addresses homogeneous
macroscopic systems in the thermodynamic limit. These are
traditionally treated in canonical ensembles controlled by
intensive temperature $T$, chemical potential $\mu$ and/or
pressure $P$. In the canonical ensemble the heat capacity is given
by the fluctuation of the energy $<\!\!(\delta E)^2\!\!>/T^2\ge
0$.

As in astro-physics the heat capacity is often negative it is immediately
clear that astro-physical systems are not in the canonical ensemble. This
was often considered as a paradoxical feature of the statistics of
self-gravitating systems. Here we will show that this is not a mistake of
equilibrium statistics when applied to self-gravitating systems but is a
generic feature of statistical mechanics of any many-body systems at phase
separation, independently of the range of the interactions,
ref.\cite{gross219}.

As the original motivation of thermodynamics was the understanding of
boiling water in steam-engines, this points to a basic misconception of
conventional canonical thermo-statistics. As additional benefit of our
reformulation of the basics of statistical mechanics by microcanonical
statistics there is a rather simple interpretation of entropy, the
characteristic entity of thermodynamics.
\section{What is entropy?\label{entropy}}
Boltzmann, ref.\cite{boltzmann1877},  defined the entropy of an isolated
system  in terms of the sum of all possible configurations, $W$, which the
system can assume consistent with its constraints of given energy, volume,
and further conserved constraints:\begin{equation}
\fbox{\fbox{\vecbm{S=k*lnW}}}\label{boltzmann0}
\end{equation}as written on Boltzmann's tomb-stone, with
\begin{equation}
W(E,N,V)=
\int{\frac{d^{3N}\nvec{p}\;d^{3N}\nvec{q}}{N!(2\pi\hbar)^{3N}}
\epsilon_0\;\delta(E-H\{\nvec{q},\nvec{p}\})}\label{boltzmann}
\end{equation} in semi-classical approximation. $E$ is the total energy, $N$ is
the number of particles and $V$ the volume. Or, more appropriate
for a finite quantum-mechanical system:
\begin{equation} W(E,N,V)= Tr[\mathcal{P}_E]\label{quantumS} =
\sum{\scriptscriptstyle\begin{array}{ll}\mbox{all eigenstates n of H with
given N,$V$,}\\\mbox{and } E<E_n\le E+\epsilon_0\nonumber
\end{array}}
\end{equation}
and $\epsilon_0\approx$ the macroscopic energy resolution.  This
is still up to day the deepest, most fundamental, and most simple
definition of entropy. {\em There is no need of the thermodynamic
limit, no need of concavity, extensivity, and homogeneity}.
Schr\"odinger was wrong saying that microcanonical statistics is
only good for diluted systems, ref.\cite{schroedinger46}. It may
very well also address the solid-liquid transition
ref.\cite{gross213} and even self-gravitating systems as we will
demonstrate in this article. In its semi-classical approximation,
eq.(\ref{boltzmann}), $W(E,N,V,\cdots)$ simply measures the area
of the sub-manifold of points in the $6N$-dimensional phase-space
($\Gamma$-space) with prescribed energy $E$, particle number $N$,
volume $V$, and some other time invariant constraints which are
here suppressed for simplicity. Because it was Planck who coined
it in this mathematical form, I will call it the Boltzmann-Planck
principle.

The Boltzmann-Planck formula has a simple but deep physical interpretation:
$W$ or $S$  measure our ignorance about the complete set of initial values
for all $6N$ microscopic degrees of freedom which are needed to specify the
$N$-body system unambiguously, ref.\cite{kilpatrick67}. To have complete
knowledge of the system we would need to know [within its semiclassical
approximation (\ref{boltzmann})] the initial positions and velocities of
all $N$ particles in the system, which means we would need to know a total
of $6N$ values. Then $W$ would be equal to one and the entropy, $S$, would
be zero. However, we usually only know the value of a few parameters that
are conserved or change slowly with time, such as the energy, number of
particles, volume and so on. We generally know very little about the
positions and velocities of the particles. The manifold of all these points
in the $6N$-dim. phase space, consistent with the given conserved
macroscopic constraints of $E,N,V,\cdots$, is the microcanonical ensemble,
which has a well-defined geometrical size $W$ and, by equation
(\ref{boltzmann0}), a non-vanishing entropy, $S(E,N,V,\cdots)$. The
dependence of $S(E,N,V,\cdots)$ on its arguments determines completely
thermostatics and equilibrium thermodynamics.

Clearly, Hamiltonian (Liouvillean) dynamics of the system cannot create the
missing information about the initial values - i.e. the entropy
$S(E,N,V,\cdots)$ cannot decrease. As has been further worked out in
ref.\cite{gross183} and more recently in ref.\cite{gross207} the inherent
finite resolution of the macroscopic description implies an increase of $W$
or $S$ with time when an external constraint is relaxed, c.f.chapter
\ref{second}. Such is a statement of the second law of thermodynamics,
ref.\cite{prigogine71}, which requires that the {\em internal} production
of entropy be positive or zero for every spontaneous process. Analysis of
the consequences of the second law by the microcanonical ensemble is
appropriate because, in an isolated system (which is the one relevant for
the microcanonical ensemble), the changes in total entropy must represent
the {\em internal} production of entropy, see above, and there are no
additional uncontrolled fluctuating energy exchanges with the environment.
\section{No phase separation (no boiling water) without convex non-extensive
{\boldmath$S(E)$}.}\label{convex}
The weight $e^{S(E)-E/T}$ of configurations with energy E in the
definition of the canonical partition sum
\begin{equation}
Z(T)=\int_0^\infty{e^{S(E)-E/T}dE}\label{canonicweight}
\end{equation} becomes here {\em bimodal}, at the transition temperature it has
two peaks, the liquid and the gas configurations which are
separated in energy by the latent heat. Consequently $S(E)$ must
be convex (like $y=x^2$) and the weight in (\ref{canonicweight})
has a minimum between the two pure phases. Of course, the minimum
can only be seen in the microcanonical ensemble where the energy
is controlled and its fluctuations forbidden. Otherwise, the
system would fluctuate between the two pure phases (inter-phase
fluctuation) by an, for macroscopic systems even macroscopic,
energy $\Delta E\sim E_{lat}\propto N$ of the order of the latent
heat. I.e. {\em the convexity of $S(E)$ is the generic signal of a
phase transition of first order and of phase-separation,
ref.\cite{gross174}}. This applies also to macroscopic systems
coupled by short or long-range interactions. Of course, even there
phase-separation exists and there is a minimum (convexity) in
$S(E)$ and, consequently, negative heat capacity\cite{gross219}.
\begin{figure}[h]
\resizebox{12cm}{!}{ {\includegraphics[bb =90 8 464
643,width=12cm,angle=-90,clip=true]{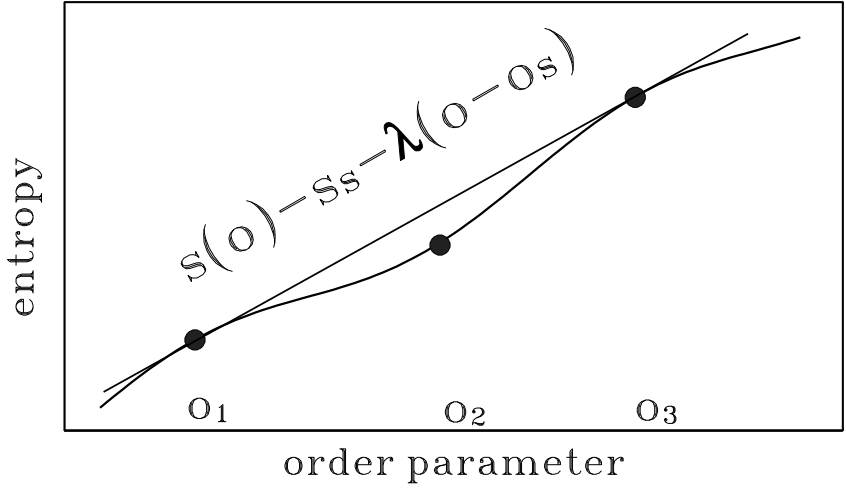}}} \caption{Convex intruder
in the entropy as function of an conserved order parameter. Gibbs double
tangent rule $\equiv$ concave hull. The derivative gives the Maxwell line
in $T(E)$. {\em Three} values of the order parameter correspond to the {\em
single} intensive ``temperature'' (inverse slope $T=1/\lambda$). At $o_1$
the system is in the ``liquid'' phase, at $o_3$ in its ``gas'' phase. The
difference $o_3-o_1$ is the latent ``heat'' and of macroscopic size
$\propto N$.} \label{fig1}
\end{figure}

It must be emphasized: In conventional canonical statistics phase-separated
systems are treated by artificial constructions of two coexisting simple
configurations of each phase like a single drop with vapor around
(Wulff-construction). But this is not a common ensemble of various droplets
with its inter-phase fluctuations leading e.g. to a negative heat capacity.
Such macroscopic energy fluctuations and the resulting negative heat
capacity are already early discussed in high-energy physics by Carlitz,
ref.\cite{carlitz72}.

\section{Application to astrophysics}
Classical Boltzmann-Planck microcanonical statistics applies very well to
the equilibrium of self-gravitating systems provided we put the system into
a box (ignoring slow evaporation) and use a hard core in the interaction at
short distances where anyhow non-gravitational physics like hydrogen
burning dominates gravity. Padmanabhan in his stimulating review
ref.\cite{padmanabhan02} discussed this point in detail.

We simplify the original many-body problem by the mean-field approximation,
details in ref\cite{gross207}. Here the (certainly also important)
many-body correlations are ignored. The entropy is then a function of the
one-body densities  $\rho(\vecbm{r})$. $S(E,\rho)$ is maximized and a
closed equation for the optimal one-body density distributions obtained. To
get this result and to get finite $\rho(\vecbm{r})$ the short-distance
cut-off and a hard core in the two-body interaction $U(\vecbm{r-r'})$ is
essential. The resulting mean-field theory therefore does {\em not} lead to
an isothermal sphere. The latter is a solution of Poisson's equation
(ref.\cite{padmanabhan02}) with a strict $1/r$ interaction.

Moreover, for equilibrium one does not need any exotic q-deformed
statistics, ref.\cite{gross203,tsallis04}. The error of
ref.\cite{tsallis04} is the use of a generalized variant of Boltzmann-Gibbs
{\em canonical} ensemble instead of the basic microcanonical one, a point
on which Tsallis seems to agree in his discussion in ref.\cite{tsallis04},
see also ref.\cite{gross207}.

In fig. \ref{cover} we show the equilibrium distributions of
self-gravitating systems under large angular-momen\-tum in the
mean-field approximation:
\begin{figure}[h]\resizebox{12cm}{!}
{\includegraphics*[bb = 93 401 519 592,
angle=-0, width=12cm, clip=tru]{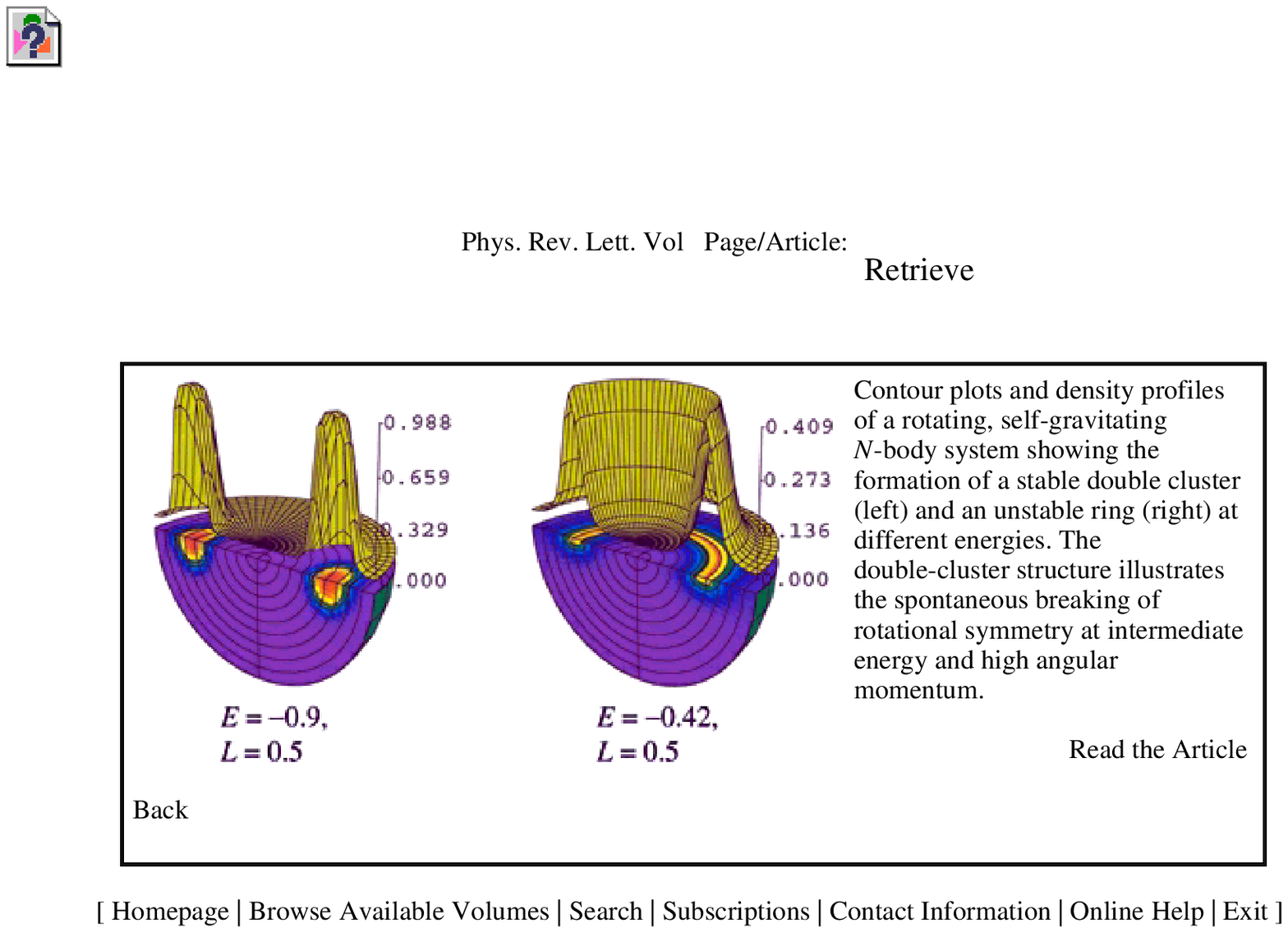}}
\caption{Rotating multi-star-systems at equilibrium. The left shows a
rotating double-star system in the DC phase, c.f. fig.\ref{phased}. This is
an inhomogeneous phase, analog to nuclear multifragmentation. The right is
one of the many different, unstable, configurations existing in the mixed
phase with negative heat capacity. Here the system fluctuates between such
ring systems, systems of stars rotating around a central star but also
mono-stars and eventual gas. This region is very interesting but must still
be more investigated. [Cover page of Phys.Rev.Lett. vol 89, (July
2002)]\label{cover}}
\end{figure}

The necessity of using ``extensive'' instead of ``intensive''
control parameters is explicit in astrophysical problems. E.g.:
for the description of rotating stars one conventionally works at
a given temperature  and fixed angular velocity $\Omega$ c.f.
ref.\cite{chavanis03}. Of course in reality there is neither a
heat bath nor a rotating disk. Moreover, the latter scenario is
fundamentally wrong as at the periphery of the disk the rotational
velocity may even become larger than velocity of light.
Non-extensive systems like astro-physical ones do not allow a
``field-theoretical'' description controlled by intensive fields !
\begin{figure}[h]\resizebox{12cm}{!}
{\includegraphics[bb =0 0 511
353,width=12cm,angle=0,clip=true]{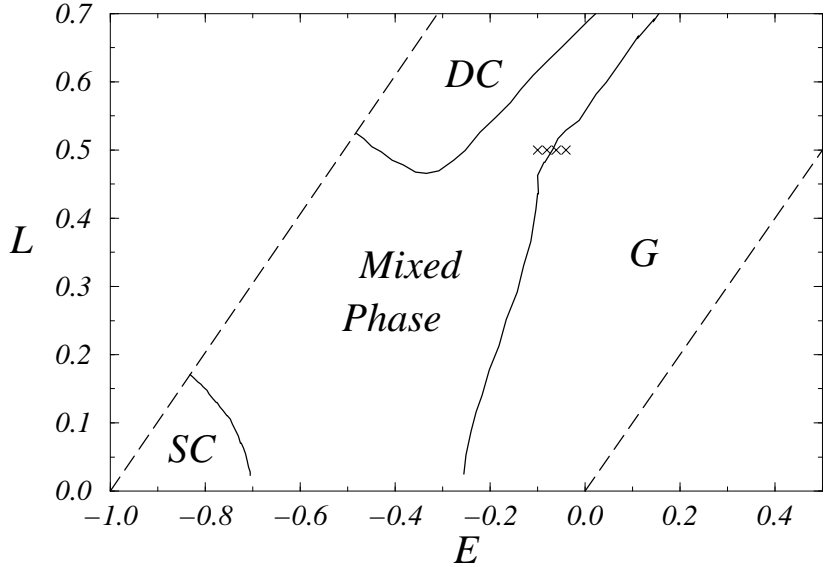}}
 \caption{Phase diagram of rotating self-gravitating systems
in the energy-angular-momentum $(E,L)$-plane. DC: region of double-stars,
G: gas phase, SC: single stars. In the mixed region one finds various
exotic configurations like ring-systems in coexistence with gas, double
stars or single stars. In this region of phase-separation the heat capacity
is negative and the entropy is convex. The dashed lines $E-L=-1$ (left) and
$E=L$ (right) delimit the region where systematic calculations were carried
out. At a few points outside of this strip some calculations like the left
of fig.(\ref{cover}) were also done. \label{phased}}
\end{figure}

E.g. configurations with a maximum of random energy on a rotating
disk, i.e. at fixed rotational velocity $\Omega$ (which is
intensive):
\begin{equation}
E_{random}=E-\frac{\Theta\Omega^2}{2} -E_{pot}
\end{equation} and consequently with the largest entropy are the ones
with smallest moment of inertia $\Theta$, compact single stars.
Just the opposite happens when the angular-momentum $L$
(extensive) and not the angular velocity $\Omega$ is
fixed:\begin{equation} E_{random}=E-\frac{L^2}{2 \Theta} -E_{pot}.
\end{equation}Then configurations with large moment of inertia are
maximizing the phase space and the entropy. I.e. eventually double
or multi stars are produced, as observed in reality.

In figure \ref{phased} one clearly sees the rich and realistic
microcanonical phase-diagram of a rotating gravitating system controlled by
the conserved ``extensive'' parameters energy and angular-momentum,
ref.\cite{gross187}.

\section{On the Second Law \label{second}}
Also the Second Law becomes significantly more transparent within
microcanonical statistics:

Many formulations of the second law exist. The understanding of
entropy is sometimes obscured by frequent use of the
Boltzmann-Gibbs canonical ensemble, and thermodynamic limit; and
the relationship of entropy to the second law is often beset with
confusion between external transfers of entropy and its internal
production. Perhaps the clearest statement is by Prigogine,
ref.\cite{prigogine71}: He distinguishes between internal and
external entropy transfer. Whereas the transfer $d_{ext}S=dQ/T$ to
an external heat bath can be positive, negative, or zero, the
internal generation of entropy $d_{int}S$ within the system is
necessarily $d_{int}S\ge 0$. This is the sharpest formulation of
the second law. $d_{int}S$ is best seen in the microcanonical
ensemble, as there are no couplings to any heat bath and no
sometimes uncontrolled exchanges of energy with it.

The second law is then most rigorously given by equation (\ref{secondlaw}):
\begin{eqnarray}
dS&=&d_{ext}S+d_{int}S\nonumber\\ d_{ext}S&=&\mbox{positive, negative, or
zero}\nonumber\\d_{int}S &\geq& 0\label{secondlaw}
\end{eqnarray}

\begin{figure}[h]\resizebox{12cm}{!}
{
\begin{minipage}[t]{6cm}
\begin{center}$V_a$\hspace{2cm}$V_b$\end{center}
\vspace*{0.2cm}
\includegraphics*[bb = 0 0 404 404, angle=-0, width=5.7cm,
clip=true]{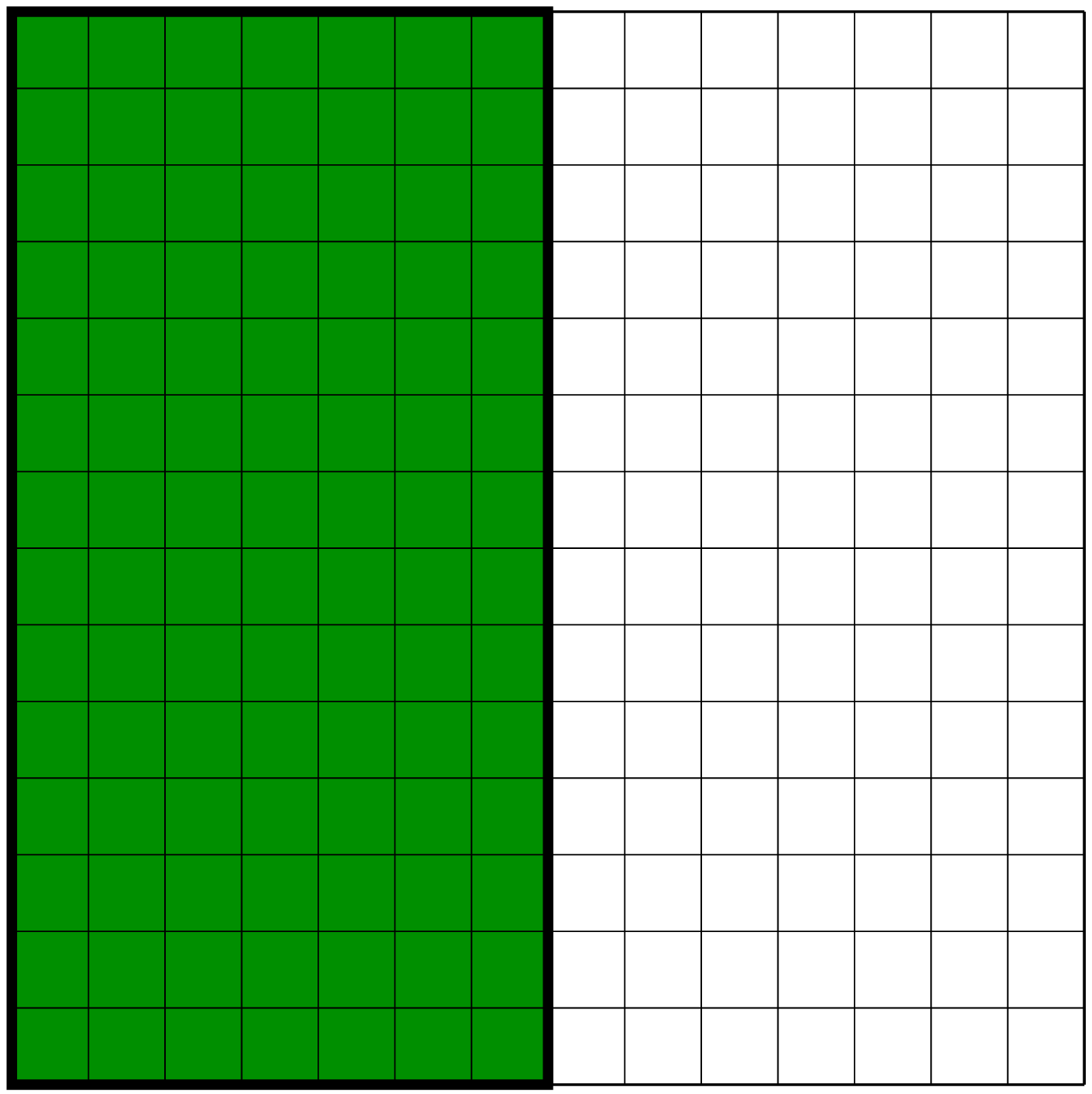}\begin{center}$t<t_0$\end{center}
\end{minipage}\lora\begin{minipage}[t]{6cm}
\begin{center}$V_a+V_b$\end{center}
\includegraphics*[bb = 0 0 428 428, angle=-0, width=6cm,
clip=true]{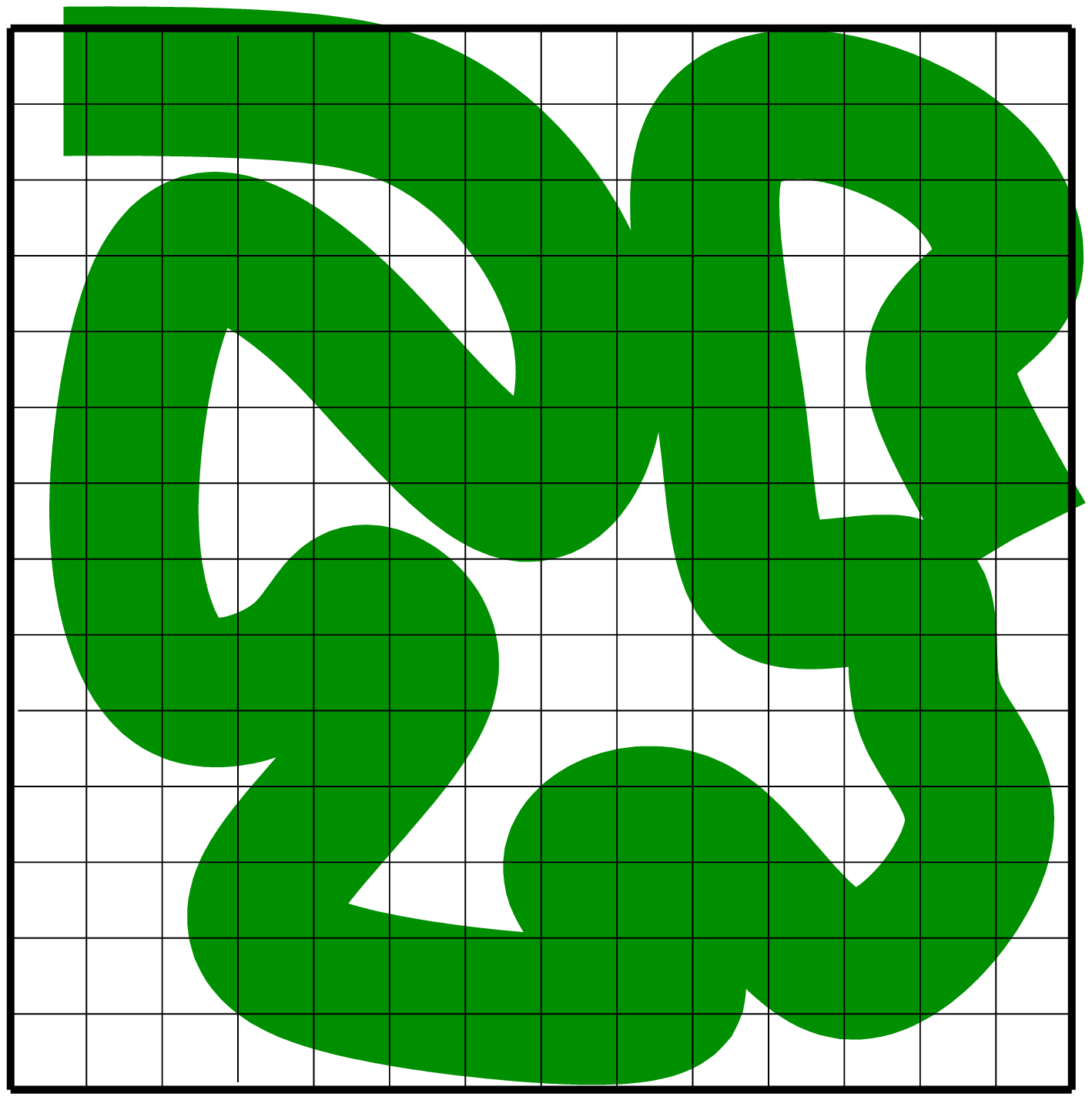}\begin{center}$t>t_0$\end{center}
\end{minipage}}
\caption{The sudden expansion in phase-space of an equilibrized system left
into a larger phase-space at $t>t_0$ right. When at $t_0$ the wall closing
volume $V_a$ is suddenly removed, the compact set $M(t<t_0)$, left side,
develops into an increasingly folded ``spaghetti''-like, though still
compact, distribution in phase-space with rising time $t>t_0$, the Gibbs'
"ink-lines". The right figure shows only the early form of the
distribution. At much larger times it will become more and more fractal and
finally dense in the new phase space.  The grid illustrates the boxes of
the box-counting method. All boxes with size $\delta^{6N}$ which overlap
with $M(t)$ are counted by $N_\delta$ in eq.(\ref{boxvol})\label{box}}
\end{figure}

A geometric visualization how an equilibrium distribution in the $6N$-dim
phase space expands into a (fat)-fractal (Gibbs ink lines) after a
constraint, here the wall between the phase-space volumes between $V_1$ and
$V_2$ is suddenly released is given by figure (\ref{box}).

The Liouville theorem tells us that the (Riemannian-) size of it
as given by eq.(\ref{boltzmann}) does not change. Keeping in mind
that due to the redundant information about the system (c.f.
chapter \ref{entropy}) it is in general impossible to distinguish
a point inside the fractal from a neighboring point outside.
Therefore the proper way to ``measure'' the size of the manifold
is by its box-counting ``measure''. I.e. the phase-space integrals
in eq.(\ref{boltzmann}) or eq.(\ref{quantumS}) have to be
calculated by the {\em box-counting volume} $\mbox{vol}_{box}{\bf
M}$ as is explained in fig.(\ref{box}) (details in
refs.\cite{gross183,gross207}):
\begin{equation}
\mbox{vol}_{box}[{\bf M}(E,N,t\gg t_0)]:=\underbar{$\lim$}_{\delta\to 0}
\delta^{6N} N_\delta[{\bf{M}}(E,N,t\gg t_0)]\label{boxvol}
\end{equation}
which is  $\ge\mbox{vol}_{box}[{\bf M}(E,N,t< t_0)]$. That is the Second
Law.

\section{Further challenges} The story of applying microcanonical
statistics to self-gravitating systems and eventually to cosmology is far
open. There are many questions to answer, e.g.:

There is the problem of statistics in an expanding universe as discussed by
Padmanabhan ref.\cite{padmanabhan02}, the question of defining
Boltzmann-Planck statistics in general relativity and connected the problem
of avoiding field-theoretical descriptions with intensive statistical
control-parameters or using them despite the fact that first order
transitions demand extensive control parameters as discussed here. The
present author, coming from far away from nuclear and statistical physics,
is open for any helpful suggestion. An interdisciplinary meeting like this
NBS2005 is here very important. I like to thank Francoise Combes and Raoul
Robert for this great opportunity.
\section{Conclusion}
Conventional canonical statistics is insufficient to handle the original
goal of Thermodynamics, phase separations with their characteristic
inter-phase fluctuations. Here the microcanonical ensemble is much richer
and leads to {\em negative specific heat}. This is thus {\em not} a
characteristics of self-gravitating systems. Phase transitions can be found
{\em without the thermodynamic limit, without homogeneity, concavity, and
extensivity}. Microcanonical statistics opens to a much wider class of
applications as e.g. astrophysical systems. Moreover it gives a simple
formulation and proof of the second law.


\end{document}